\documentclass[pra,reprint,showpacs,floatfix]{revtex4-2}
\usepackage{amsmath,amsthm,amssymb}
\usepackage[pdftex]{graphicx}
\usepackage{bm}

\begin{document}

\title{Exact expressions for the height of the interatomic step in the exchange-correlation potential from the derivative discontinuity of the energy}

\date{\today}

\author{M.\ J.\ P.\ Hodgson}
\email[Personal email: ]{matthew.j.hodgson@durham.ac.uk}
\homepage[Personal webpage: ]{\url{http://www-users.york.ac.uk/~mjph501/}}
\affiliation{Department of Physics, Durham University, South Road, Durham, DH1 3LE, United Kingdom}
\affiliation{The European Theoretical Spectroscopy Facility}

\begin{abstract}
Popular approximations to the exchange-correlation (xc) energy of density functional theory do not yield the spatial `step' structures in the exact xc potential which are necessary to describe dissociation and electron excitation with the Kohn-Sham (KS) system. Via the discontinuity in the derivative of the xc energy as a function of electron number I derive exact analytic expressions in terms of the KS single-particle energies for the height of the step in the xc potential between a variety of open- and closed-shell atoms within stretched molecules. 
\end{abstract}

\maketitle

\section{Introduction}

The unparalleled success of Kohn-Sham (KS) \cite{PhysRev.140.A1133} density functional theory (DFT) \cite{PhysRev.136.B864} within solid state physics \cite{burke2012perspective,jain2016computational,verma2020status} is not experienced within quantum chemistry. DFT's popularity stems from the accuracy and computational efficiency of commonly used approximations to the exchange-correlation (xc) energy, such as the local density \cite{PhysRev.140.A1133} and the generalized gradient \cite{PhysRevLett.77.3865} approximation. However, these popular approximations are unreliable for molecules with strong electron localization \cite{mori2008localization,graziano2017quantum,laplaza2019localizing}. 

An accurate ground-state KS potential is required for determining ground-state properties, such as the total energy \cite{medvedev2017density}, as well as the optical absorption spectra \cite{PhysRevB.101.115109,PhysRevB.99.161102}, charge-transfer energies \cite{kimber2020toward,doi:10.1021/acs.jctc.0c01093,kocak2021charge} and electron real-time dynamics \cite{burke2005time}. As existing approximations within KS theory are unreliable, computationally demanding hybrid density functionals \cite{perdew1996rationale,seidl1996generalized} are employed to calculate these properties \cite{salzner1997design,heyd2003hybrid,isegawa2012performance,pastore2013modeling,pyzer2015high,hait2016prediction,kummel2017charge,PhysRevMaterials.2.040801,PhysRevB.99.045129,verma2020status}.

The exact xc potential has a strong dependence on the electron density \textit{everywhere} in the system which is challenging to capture within an approximation. Within the KS system the KS potential ensures that the non-interacting single-particle electron density equals the many-body density. For systems in which there is a large distance, $d$, between its subsystems, e.g., the disassociated atoms of a molecule or a donor-acceptor pair, the exact xc potential can be discontinuous as a function of space. Such features are termed spatial `steps'. They occur at points in the density where the `local effective ionization potential' (LEIP) \cite{PhysRevB.93.155146} changes, for example at the intersection of two subsystems. The step can be present in the xc potential in the limit that the subsystems are infinitely separated \cite{NATO85_AvB,PhysRevA.94.052506,PhysRevB.93.155146} and depend on properties of each subsystem; see Eq.~\eqref{Eq:AvB}. 

For diatomic molecules the step that occurs \textit{between} the atoms can be crucial to obtain the correct electron occupation on each atom throughout the ground- and excited-state KS system \cite{PhysRevB.93.155146,doi:10.1021/acs.jctc.0c01093}. Common approximations based on the local density that omit this `interatomic' step yield spurious fractional electron numbers on the disassociated atoms \cite{:/content/aip/journal/jcp/125/19/10.1063/1.2387954,PhysRevA.83.062512,hofmann2012integer,armiento2013orbital,mori2008localization,wang2020orbital}, termed the delocalization error \cite{cohen2012challenges,hellgren2019strong}, which leads to an unreliable prediction of the bonding length and dissociation energies \cite{:/content/aip/journal/jcp/131/22/10.1063/1.3271392}. Despite employing a spatially nonlocal potential, the \textit{exact} multiplicative potential of hybrid density functionals \cite{seidl1996generalized} can also require an interatomic step to accurately describe atomic dissociation \cite{PhysRevA.101.032502}.

For a system that consists of two separated open-shell atoms the interatomic step height, $S$, is given by Almbladh and von Barth's expression \cite{NATO85_AvB}
\begin{equation} \label{Eq:AvB}
    S = I_\mathrm{L} - I_\mathrm{R},
\end{equation}
where $I_\mathrm{L}$ is the ionization energy of the left atom and $I_\mathrm{R}$ the right. An analytic expression for the step height in terms of the atomic KS quantities has not been formally derived for a system that contains any closed-shell atoms. 

The electron number of a finite system can vary if it is in contact with an electron reservoir, the chemical potential of which can be adjusted. The exact xc potential, $v_\mathrm{xc}(x)$, experiences a discontinuous \textit{uniform} shift by a constant, $\Delta^{N_0}$, when the number of electrons, $N$, within the system infinitesimally surpasses an integer \cite{perdew1982density}, $N_0$. $\Delta^{N_0}$ corresponds to the discontinuity in the derivative of the xc energy with respect to $N$ at $N=N_0$. The magnitude of the shift is given by
\begin{align} \label{Eq:DD}
\Delta^{N_0} &= \lim_{\delta \rightarrow 0} \left ( \left. v_\mathrm{xc} (x) \right |_{N=N_0+\delta} - \left . v_\mathrm{xc} (x) \right |_{N=N_0-\delta} \right ) \nonumber \\
&= I^{N_0} - A^{N_0} - \left( \varepsilon^{N_0}_{N_0+1}-\varepsilon^{N_0}_{N_0} \right),
\end{align}
where $\delta$ is the amount of additional electron ($0 \leq \delta \leq 1$), $I^{N_0}$ is the ionization energy, $A^{N_0}$ is the electron affinity, $\varepsilon^{N_0}_{N_0}$ is KS energy of the highest occupied molecular orbital (HOMO) and $\varepsilon^{N_0}_{N_0+1}$ is the KS energy of the lowest unoccupied molecular orbital (LUMO), all of the $N_0$-electron system.

This xc `derivative discontinuity' in principle yields the fundamental gap (the ionization energy minus the electron affinity) from the KS system \cite{perdew1982density,perdew1983physical,kraisler2014fundamental}; see Eq.~(\ref{Eq:DD}). In practice, common density-functional approximations do not capture this discontinuity \cite{:/content/aip/journal/jcp/125/19/10.1063/1.2387954,sousa2007general,:/content/aip/journal/jcp/129/12/10.1063/1.2987202,cohen2008insights,mori2014derivative,mosquera2014derivative}.

In Refs.~\onlinecite{hodgson2017interatomic} and \onlinecite{doi:10.1021/acs.jctc.0c01093} the relationship between the interatomic step and the discontinuity in the derivative of the xc energy was demonstrated in the context of charge transfer within a diatomic molecule. 

In this article I derive analytic expressions for the interatomic step height for stretched molecules that consist of open- and closed-shell atoms via the xc derivative discontinuity. These expressions could be employed within practical density-functional calculations, for example within approximate functionals that yield step structures in the xc potential but require exact conditions for practical use \cite{PhysRevB.90.241107}. 

I employ the \texttt{iDEA} code \cite{PhysRevB.88.241102} to verify my analytic results via exactly solvable model systems. The code models few-electron systems in 1D exactly by solving the many-body Schr\"odinger equation on a real-space grid. The code employs the softened Coulomb interaction $(\left | x - x' \right | + 1)^{-1}$ as appropriate \cite{PhysRevA.72.063411}. From the exact fully-correlated, ground-state many-body wavefunction I obtain the exact ground-state many-body density, $n(x)$. I reverse-engineer the KS equations so that the KS density is equal to the many-body density to find the corresponding \textit{exact} KS potential, orbitals and energies; details of the reverse-engineering procedure are given in Ref.~\onlinecite{PhysRevB.88.241102}.

\section{Adding an infinitesimal amount of an electron to a diatomic molecule} \label{sec:infinitesimal_additional_e}

\begin{figure}[htbp]
  \centering
  \includegraphics[width=1.0\linewidth]{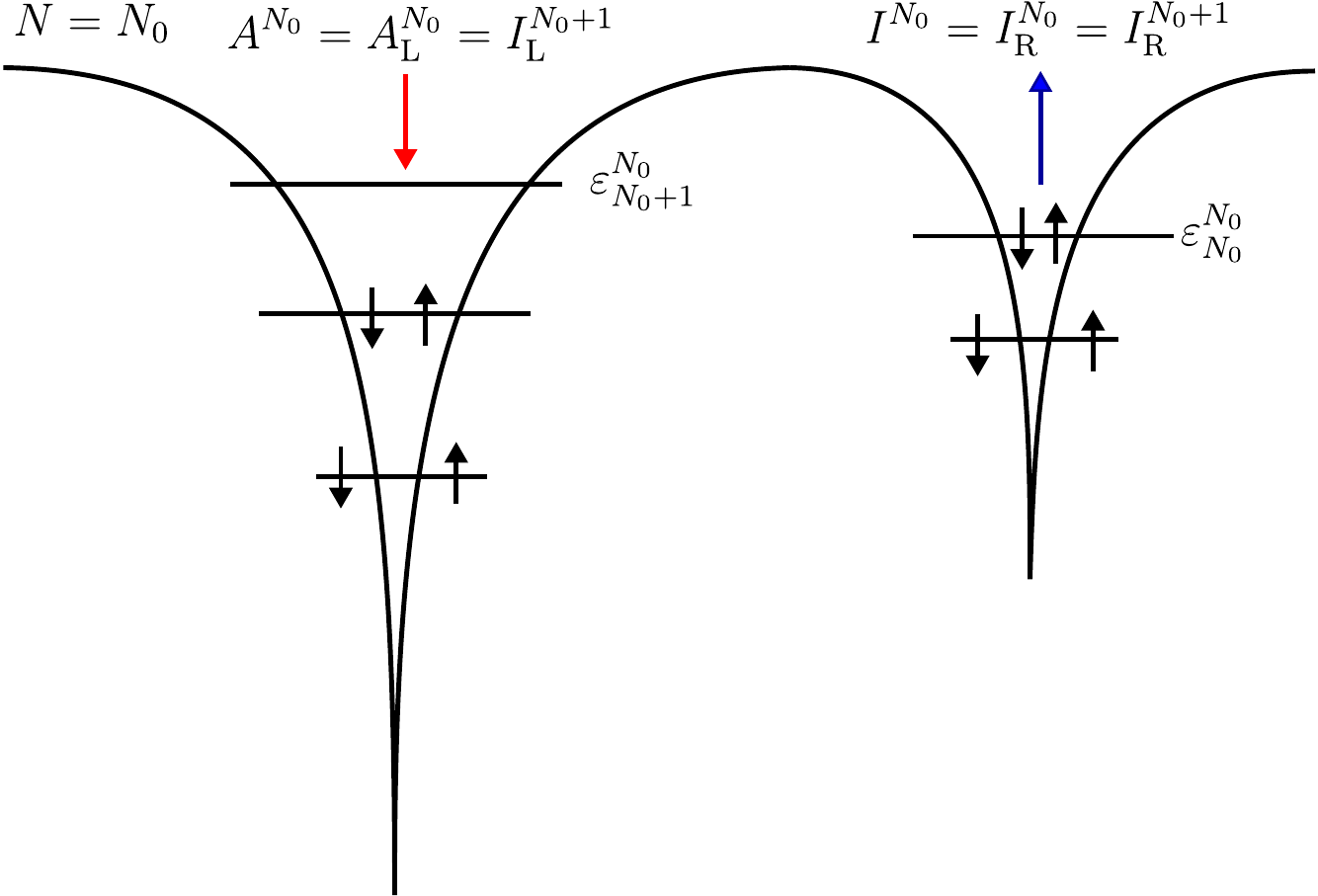}
\caption{Diagram of an $N_0$-electron stretched diatomic molecule. Both atoms are closed shell within the exact KS description. The lowest unoccupied KS orbital is localized to the left atom and the highest occupied KS orbital is localized to the right atom.}
\label{fig:first_molecule}
\end{figure}

Consider the exact KS description of an $N_0$-electron stretched diatomic molecule. The external potential generated by the left nucleus (L) corresponds to a deeper potential well than that of the right (R) and both atoms are closed shell; see Fig.~\ref{fig:first_molecule}. The HOMO is localized to the right atom and the LUMO is localized to the left atom. \textit{The atoms are so separated that adding or subtracting electrons from either atom has a negligible effect on the electron density localized to the other atom.}

An electron added to this $N_0$-electron system localizes to the left atom and ionization causes an electron localized to the right atom to escape. Hence the ionization energy of the \textit{whole} system corresponds to the ionization energy of the isolated right atom ($I^{N_0}=I^{N_0}_\mathrm{R}$) and the whole system's electron affinity is that of the isolated left atom ($A^{N_0}=A^{N_0}_\mathrm{L}$); see Fig.~\ref{fig:first_molecule}. 

According to Eq.~(\ref{Eq:DD}) the xc derivative discontinuity of the \textit{whole} $N_0$-electron system (left and right atom) is $\Delta^{N_0} = I^{N_0}_\mathrm{R} - A^{N_0}_\mathrm{L} - \left( \varepsilon^{N_0}_{N_0+1} - \varepsilon^{N_0}_{N_0} \right )$. As the exact xc potential of the $N_0$-electron system tends to zero at infinity, $v_\mathrm{xc}(\left|x\right| \rightarrow \infty) = 0$, $I^{N_0}_\mathrm{R}=-\varepsilon^{N_0}_{N_0}$ (the `IP theorem' of DFT). Therefore $\Delta^{N_0} = - A^{N_0}_\mathrm{L} - \varepsilon^{N_0}_{N_0+1}$. 

For this system the LUMO is also the left atom's (the atom to which the additional electron localizes) lowest unoccupied KS \textit{atomic} orbital. Thus the xc derivative discontinuity of the isolated left atom $\Delta^{N_0}_\mathrm{L} =- A^{N_0}_\mathrm{L} - \varepsilon^{N_0}_{N_0+1}$ (where the $N_0$ corresponds to the fact that the left atom is within the $N_0$-electron system) which is equal to $\Delta^{N_0}$ for this molecule. 

\begin{figure}[htbp]
  \centering
  \includegraphics[width=1.0\linewidth]{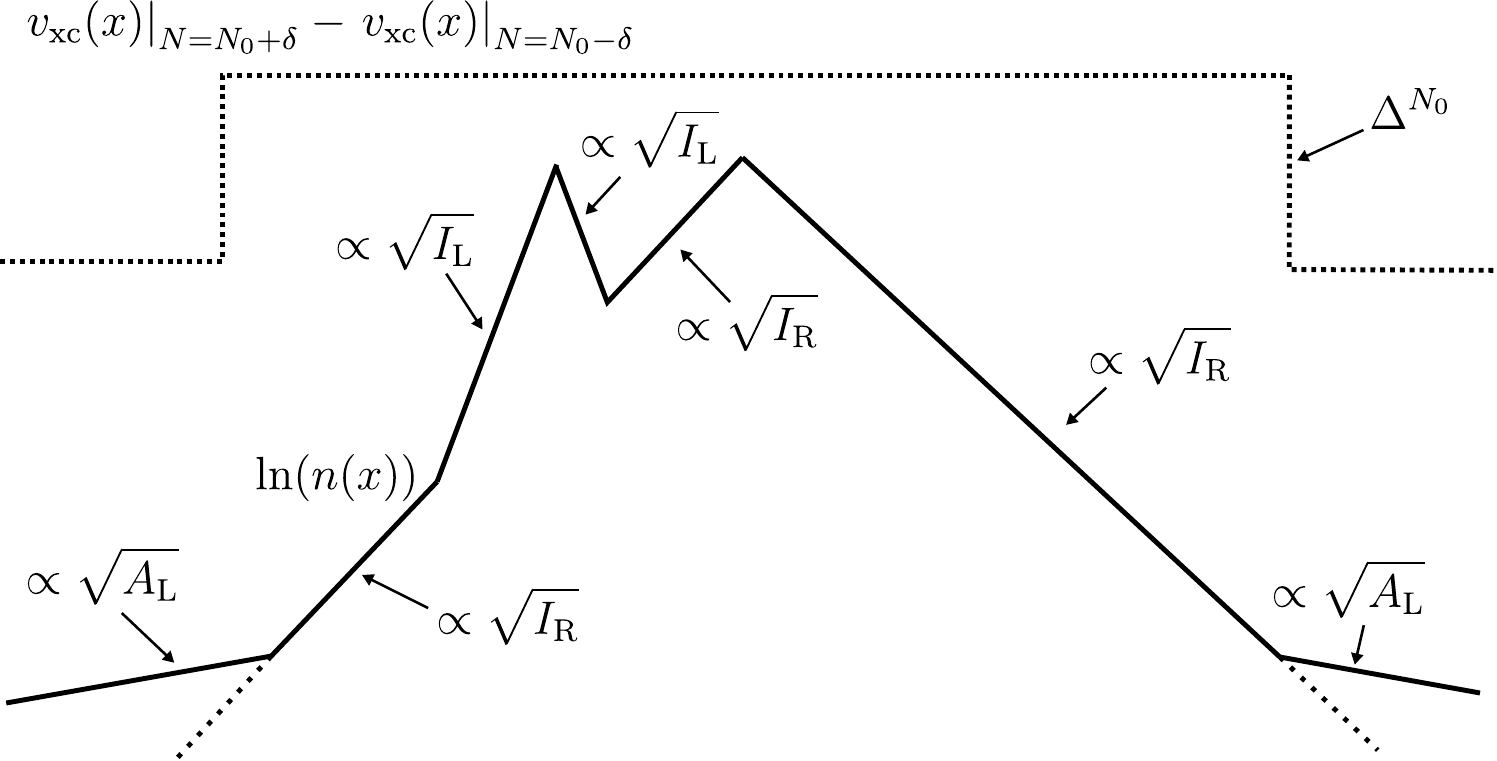}
\caption{Diagram of the ($N_0+\delta$)-electron system with a vanishingly small $\delta$. On the periphery of the system the exponential decay of the density changes from $\propto \sqrt{I^{N_0}_\mathrm{R}}$ to $\propto \sqrt{A^{N_0}_\mathrm{L}}$ which yields a step in the xc potential of height $\Delta^{N_0}$ (see text).}
\label{fig:infinitesimal_delta}
\end{figure}

Figure~\ref{fig:infinitesimal_delta} demonstrates the change in the xc potential and the natural log of the electron density when a vanishingly small amount of an electron is added to this $N_0$-electron system. The exact many-body total density of the ($N_0+\delta$)-electron system is an ensemble of the $N_0$-electron and the ($N_0+1$)-electron systems, as such 
\begin{equation} \label{Eq:ensemble}
    n^{N_0+\delta}(x) = (1-\delta)n^{N_0}(x)+\delta n^{N_0+1}(x).
\end{equation}
Far from any system the many-body density decays exponentially $\propto e^{-2\sqrt{2I}\left| x \right|}$ \cite{PhysRevA.16.1782,katriel1980asymptotic}, where $I$ is the ionization energy of the system. Thus far from the center of my stretched molecule the density decays $\propto e^{-2\sqrt{2I^{N_0}_\mathrm{R}}\left| x \right|}$ ($I^{N_0}_\mathrm{R}$ is the whole system's ionization energy). Even when $\delta$ is vanishingly small, eventually the total density decay is dominated by the decay $e^{-2\sqrt{2A^{N_0}_\mathrm{L}}\left| x \right|}$ ($A^{N_0}_\mathrm{L}$ is the whole system's electron affinity):
\begin{align} \label{Eq:ensemble_large_x}
    &n^{N_0+\delta}(\left | x \right| \rightarrow \infty ) = \nonumber \\
    &(1-\delta)Be^{-2\sqrt{2I^{N_0}_\mathrm{R}}\left| x - \frac{1}{2}d \right|}+\delta Ce^{-2\sqrt{2A^{N_0}_\mathrm{L}}\left| x+\frac{1}{2}d \right|}, 
\end{align}
where $B$ and $C$ are normalization constants. In order for the KS density to decay with the \textit{same rate} as the many-body density, on the periphery of the system, at the point in the density where the LEIP changes, steps must form in the xc potential yielding a finite `plateau' with height $\Delta v_\mathrm{xc} = \left . v_\mathrm{xc} (x) \right |_{N=N_0+\delta} - \left . v_\mathrm{xc} (x) \right |_{N=N_0-\delta}$ (where $v_\mathrm{xc}(\left| x \right | \rightarrow \infty)=0$) localized to the molecule; see Fig.~\ref{fig:infinitesimal_delta} \footnote{These steps are depicted in Fig.~\ref{fig:infinitesimal_delta} as perfectly sharp, in practice the step may have additional features, e.g., see Fig.~\ref{fig:Vary_delta_calculation}.}. Hence, the KS density is
\begin{align} \label{Eq:KSensemble_large_x}
    &n^{N_0+\delta}(\left | x \right| \rightarrow \infty ) = (1-\delta)Be^{-2\sqrt{-2\left (\varepsilon^{N_0+\delta}_{N_0} - \Delta v_\mathrm{xc}\right )}\left| x - \frac{1}{2}d \right|} \nonumber \\
    &+\delta Ce^{-2\sqrt{-2\varepsilon^{N_0+\delta}_{N_0+1}}\left| x+\frac{1}{2}d \right|}, 
\end{align}
where $\lim_{\delta \rightarrow 0} \varepsilon^{N_0+\delta}_{N_0+1}=\varepsilon^{N_0}_{N_0+1}+\Delta v_\mathrm{xc}$ and $\lim_{\delta \rightarrow 0} \varepsilon^{N_0+\delta}_{N_0} = \varepsilon^{N_0}_{N_0}+\Delta v_\mathrm{xc}$. For the KS density to decay with the same rate as the many-body density, $\varepsilon^{N_0+\delta}_{N_0} - \Delta v_\mathrm{xc} = \varepsilon^{N_0}_{N_0} +\Delta v_\mathrm{xc} - \Delta v_\mathrm{xc} =\varepsilon^{N_0}_{N_0} = -I^{N_0}_\mathrm{R}$ \textit{and} $\varepsilon^{N_0+\delta}_{N_0+1} = \varepsilon^{N_0}_{N_0+1}+\Delta v_\mathrm{xc} = -A^{N_0}_\mathrm{L}$. Thus the height of the plateau, $\Delta v_\mathrm{xc} = -A^{N_0}_\mathrm{L}-\varepsilon^{N_0}_{N_0+1}$ which is $\Delta^{N_0}$. These steps are drawn infinitely far from the system's center as $\delta \rightarrow 0$ and the plateau becomes a uniform shift in the xc potential (see Eq.~\eqref{Eq:DD}); a numerical demonstration of which is in Ref.~\onlinecite{hodgson2017interatomic}.  

\section{Increasing the amount of additional electron to one}

On the periphery of any \textit{subsystem} within a larger system the electron density decays in accordance with that subsystem's ionization energy. The decay can abruptly change, for example, when the decay of two subsystems meet. This change in the density corresponds to a change in the LEIP which in turn can yield an abrupt spatial step in the xc potential at that point \cite{PhysRevB.93.155146}. 

\begin{figure}[htbp]
  \centering
  \includegraphics[width=1.0\linewidth]{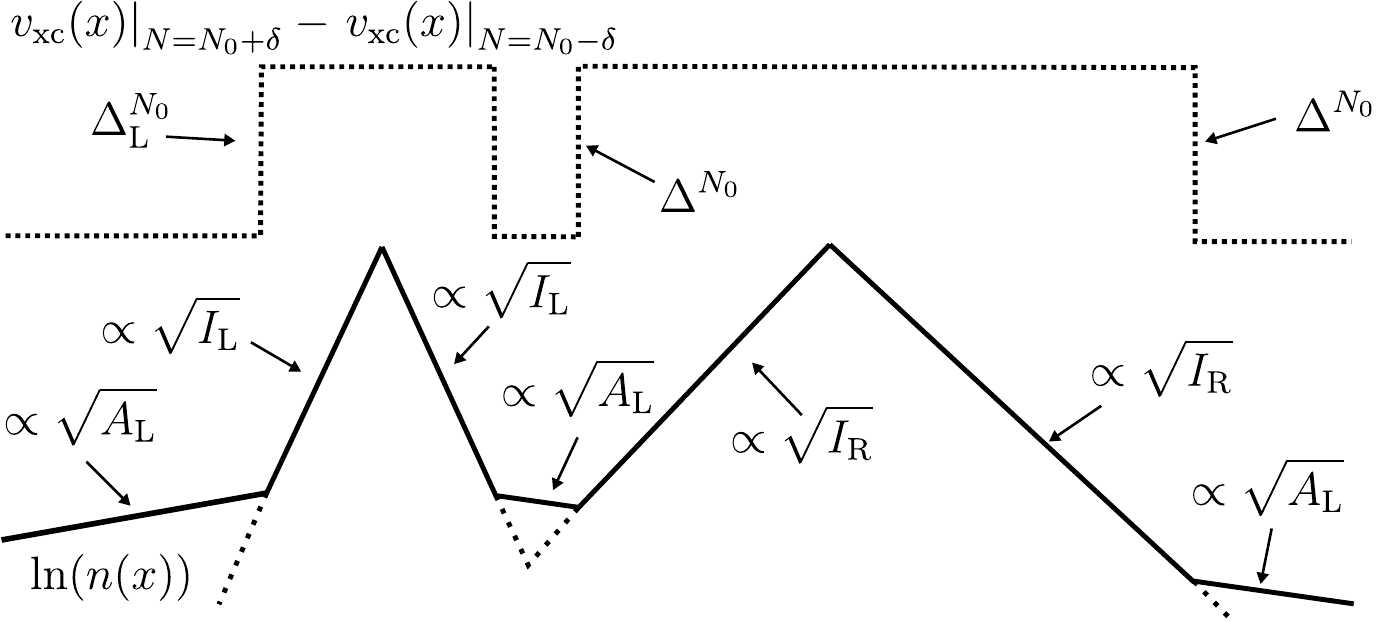}
\caption{Diagram of the ($N_0+\delta$)-electron system; $\delta$ is finite but extremely small. For the stretched molecule the LEIP changes on the periphery of the left atom resulting in a plateau with a height corresponding to the left atom's xc derivative discontinuity. The decay of the left atom then meeting the decay of the right atom yielding a second step \textit{between} the atoms. A final change in the LEIP occurs far to the right of the system. The height of the right plateau is the xc derivative discontinuity of the whole molecule in the limit that the atoms are infinitely separated (see text).}
\label{fig:small_delta}
\end{figure}

For my stretched molecule, as $\delta$ is increased slightly, the additional electron completely localizes to the left atom. Hence the left atom's local number of electrons surpasses an integer and consequently its xc potential experiences a \textit{local} shift (spatially uniform just for the left atom) by $\Delta^{N_0}_\mathrm{L}$; depicted in Fig.~\ref{fig:small_delta}. 

In Refs.~\onlinecite{hodgson2017interatomic} and \onlinecite{doi:10.1021/acs.jctc.0c01093} it was demonstrated that when an atom's local number of electrons increases by a small amount, the exponential decay of the atom's density (in my case the left atom) changes from $\propto e^{-2\sqrt{2I^{N_0}_\mathrm{L}}\left| x \right|}$ to $\propto e^{-2\sqrt{2A^{N_0}_\mathrm{L}}\left| x \right|}$ far from \textit{that atom}; e.g., in Fig.~\ref{fig:small_delta}. This causes a step to form in the xc potential around the atom, the height of which is given by that atom's xc derivative discontinuity, giving rise to the plateau in the atom's xc potential. For my stretched molecule this happens for the left atom such that the KS density decays $\propto e^{-2\sqrt{-2\varepsilon^{N_0+\delta}_{N_0+1}}\left | x \right |} = e^{-2\sqrt{-2\left (\varepsilon^{N_0}_{N_0+1}+\Delta^{N_0}_\mathrm{L}\right )}\left | x \right |} = e^{-2\sqrt{2A^{N_0}_\mathrm{L}}\left| x \right|}$. This $\sqrt{A^{N_0}_\mathrm{L}}$-decay then meets the decay of the right atom, which is $\propto e^{-2\sqrt{2I^{N_0}_\mathrm{R}}\left| x \right|}$, causing a second change in the LEIP and a second step in the xc potential between the atoms; see Fig.~\ref{fig:small_delta}.

In principle, when the atoms are far apart, the \textit{right} plateau's height does not affect the KS decay in any region which is $\propto e^{-2\sqrt{-2\varepsilon^{N_0+\delta}_{N_0+1}}\left | x \right |}$, because the highest (partially) occupied KS orbital is localized to the \textit{left} atom. Thus, the height of the right plateau (within a range \cite{PhysRevB.93.155146}) has no affect on the corresponding KS single-particle energy $\varepsilon^{N_0+\delta}_{N_0+1}$. This may lead one to assume that the height of the right plateau cannot be determined from the decay of the density. However, recall that the atoms are so far apart that the introduction of an electron to the left atom has a negligibly effect on the right atom's density, so the decays $\propto e^{-2\sqrt{2I^{N_0}_\mathrm{R}}\left| x \right|}$ and $e^{-2\sqrt{2A^{N_0}_\mathrm{L}}\left| x \right|}$ are both \textit{independent} of $\delta$. Hence the density to the right of the right atom -- given by Eq.~\eqref{Eq:ensemble_large_x} -- \textit{does not change its spatial structure} as $\delta$ is increased \footnote{The absolute position of the point in the density where the decay changes moves as $\delta$ is varied, which causes the position of the step to move accordingly. The magnitude of the density in this region also changes. However, neither of these changes affects the height of the step.}. As the KS potential is uniquely determined by the spatial structure of the density \cite{PhysRev.136.B864}, the height of the corresponding step in the xc potential at the point where the $\sqrt{I^{N_0}_\mathrm{R}}$-decay becomes the $\sqrt{A^{N_0}_\mathrm{L}}$-decay must remain fixed as a function of $\delta$ also. The step height is $\Delta^{N_0}$ in the limit that $\delta \rightarrow 0$, therefore the plateau localized to the right atom has a height of $\Delta^{N_0}$ for $0<\delta \leq 1$. It follows that the plateau's other step -- located \textit{between} the atoms -- has a height of $\Delta^{N_0}$ also; as shown in Fig.~\ref{fig:small_delta}. 

In Ref.~\onlinecite{hodgson2017interatomic} the double step structure between two separated atoms was demonstrated to occur for a 3D stretched diatomic molecule whose KS potential posses an interatomic step \textit{before} the addition of any fraction of an electron. I have shown that, in general, when a small amount of an electron is added to \textit{any} stretched molecule the atom to which the additional electron localizes experiences a shift by its own xc derivative discontinuity and the other atom's xc potential also shifts but by the xc derivative discontinuity of the \textit{whole} system \footnote{For the case of Ref.~\onlinecite{hodgson2017interatomic} the xc derivative discontinuity of the `acceptor' atom $\Delta_\mathrm{a} = - A_\mathrm{a} - \varepsilon^{N_0}_{N_0+1}$. The xc derivative discontinuity of the whole system is $\Delta = - A_\mathrm{a} - (\varepsilon^{N_0}_{N_0+1}+S) = \Delta_\mathrm{a} -S$ (because the acceptor atom is elevated by the interatomic step of height $S$). Therefore the overall step height between the atoms after the additional of a small fraction of an electron, $\Delta_\mathrm{a}-\Delta$, remains the step height $S$.}. 

\begin{figure}[htbp]
  \centering
  \includegraphics[width=1.0\linewidth]{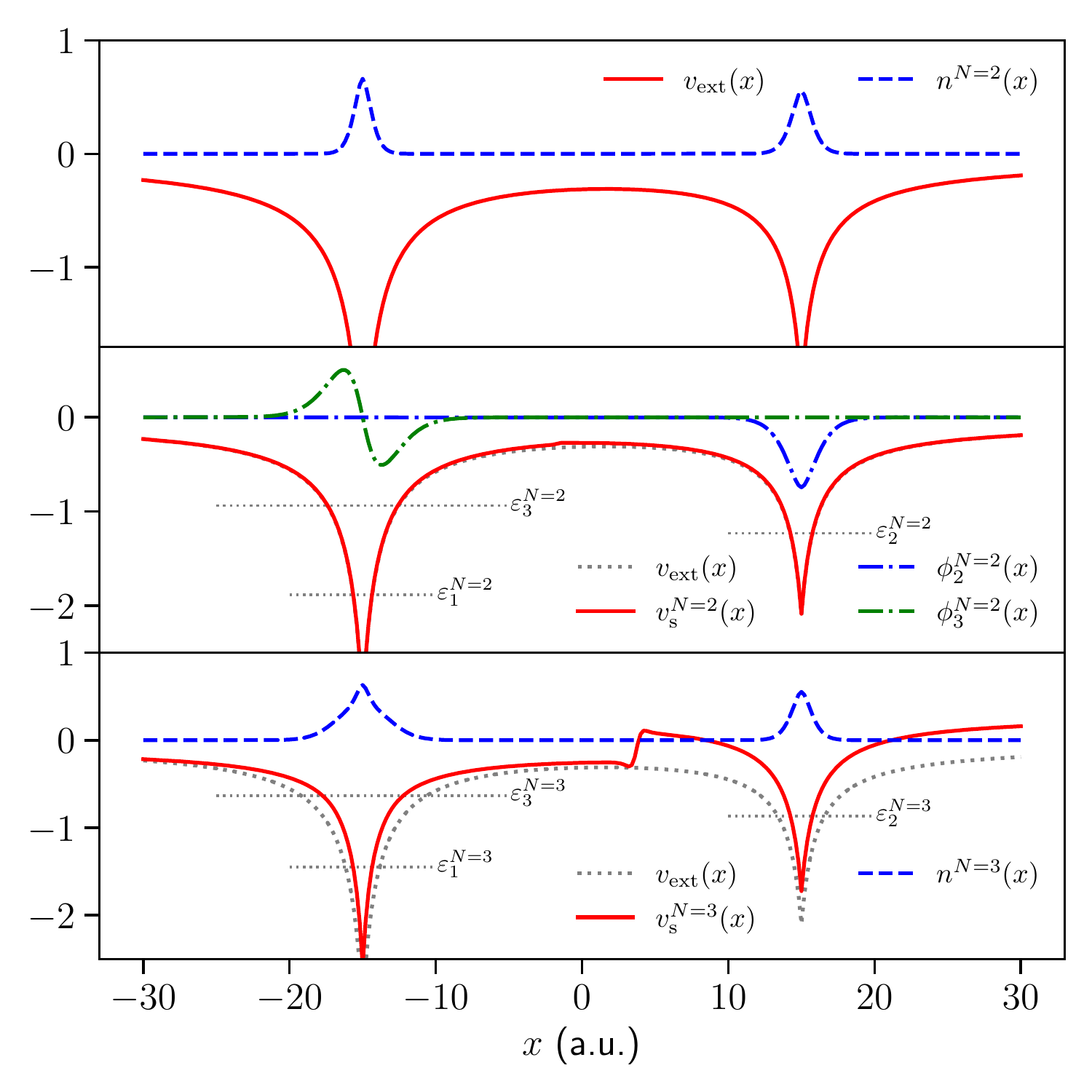}
\caption{(Top) The 2-electron density and external potential potential. (Middle) The lowest unoccupied KS orbital, $\phi^{N=2}_3(x)$, and the highest occupied KS orbital, $\phi^{N=2}_2(x)$, of the 2-electron system. The corresponding single-particle energy levels are shown in gray. (Bottom) The many-body density of three same-spin electrons and the corresponding exact KS potential. A step is present which elevates the xc potential in the vicinity of the right well the height of which is given by the xc derivative discontinuity of the 2-electron molecule.}
\label{fig:First_molecular_calculation}
\end{figure}

\begin{figure}[htbp]
  \centering
  \includegraphics[width=1.0\linewidth]{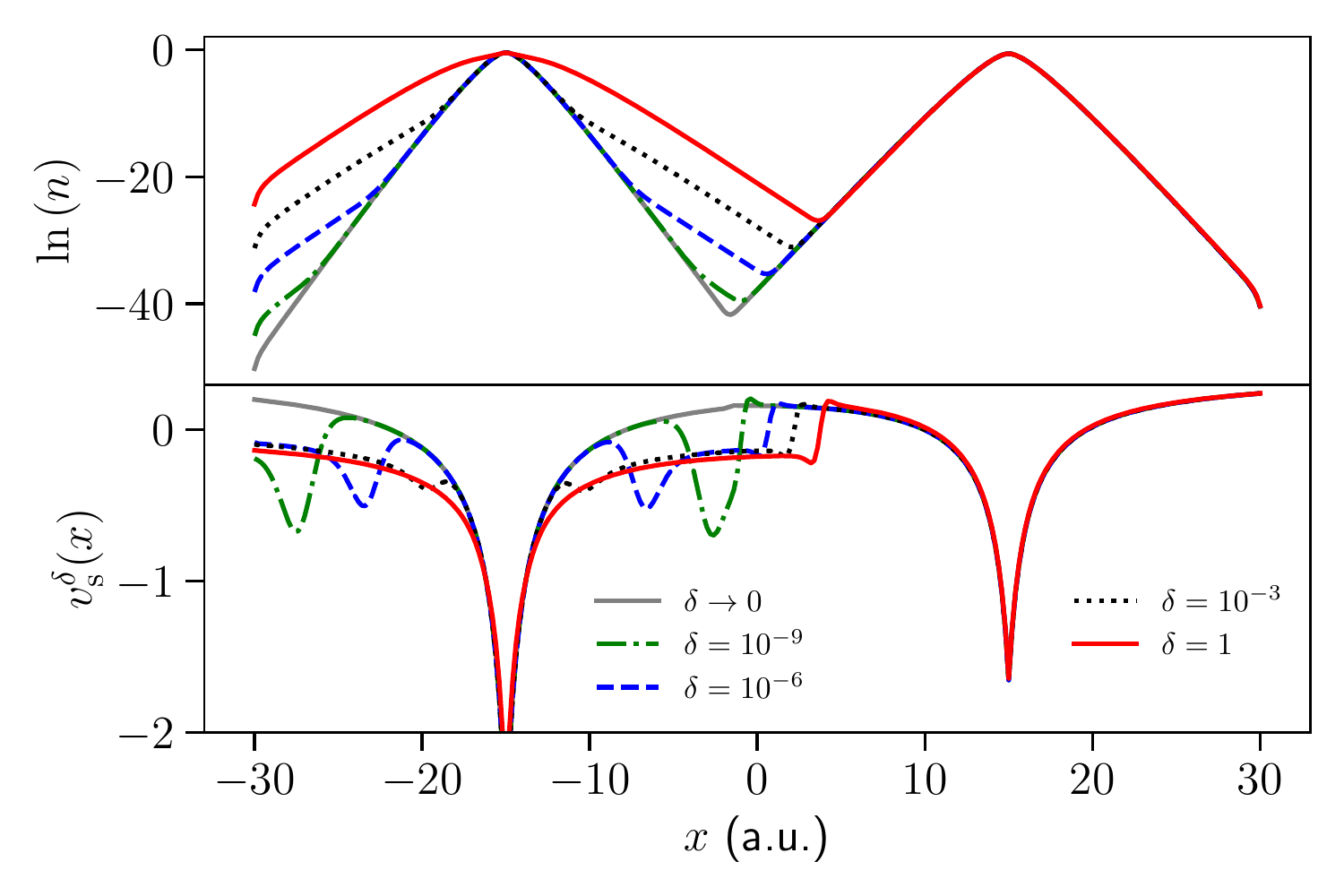}
\caption{(Top) The natural log of the density with $2+\delta$ same-spin electrons. (Bottom) The corresponding exact KS potential. As $\delta \rightarrow 1$, the additional fraction of an electron localizes to the left atom which gives rise to a plateau in its xc potential corresponding to the left atom's xc derivative discontinuity. $\Delta^{N=2}$ (the xc derivative discontinuity of the whole molecule) becomes a finite plateau localized to the right atom and remains approximately unchanged as $\delta \rightarrow 1$, ultimately giving rise to the interatomic step in the 3-electron system.}
\label{fig:Vary_delta_calculation}
\end{figure}

As $\delta \rightarrow 1$ the width of the plateau localized to the \textit{left} atom (with height $\Delta^{N_0}_\mathrm{L}$) shrinks as the LEIP in left atom's density moves closer to the center of the atom; see Fig.~\ref{fig:Vary_delta_calculation}. The plateau localized to the right atom is unaffected owing to the large separation of the atoms. Thus for $\delta = 1$ there is an interatomic step within the ($N_0+1$)-electron system's xc potential caused by the presence of the right plateau of height $\Delta^{N_0}$:
\begin{equation}
    \lim_{d \rightarrow \infty} S^{N_0+1} = \Delta^{N_0}; \nonumber
\end{equation}
see Fig.~\ref{fig:First_molecular_calculation} (bottom).

I model a 1D 2-electron stretched diatomic molecule ($N_0=2$), where the external potential is $v_\mathrm{ext}(x) = -2/(|x-\frac{d}{2}| + 1) - 3/(|x+\frac{d}{2}| + 1)$ and $d=30$ a.u.; see Fig.~\ref{fig:First_molecular_calculation} (top). All electrons are \textit{same-spin} so that each atom is closed shell. The system is designed such that the LUMO is localized to the left well and the HOMO the right; see Fig.~\ref{fig:First_molecular_calculation} (middle) and compare with Fig.~\ref{fig:first_molecule}. Within the 3-electron many-body system the additional electron has localizes to the left well; see Fig.~\ref{fig:First_molecular_calculation} (bottom). 

Via Eq.~\eqref{Eq:ensemble} I combine the 2-electron density and the 3-electron density in an ensemble to obtain the exact many-body density as a function of $\delta$. I reverse-engineer the KS equations with fractional occupation of the (now) HOMO, such that $n^{N=2+\delta}(x) = \left | \phi^{N=2+\delta}_1(x) \right |^2 + \left | \phi^{N=2+\delta}_2(x) \right |^2 + \delta \left | \phi^{N=2+\delta}_3(x) \right |^2$, to find the corresponding exact KS potential as a function of $\delta$; see Fig.~\ref{fig:Vary_delta_calculation}. 

For small $\delta$ the density in the region of the left atom is unchanged; there is a change in the LEIP either side of the left atom as expected; compare Fig.~\ref{fig:Vary_delta_calculation} with Fig.~\ref{fig:small_delta}. The corresponding steps in the xc potential give rise to the plateau which elevates the left atom's xc potential by the xc derivative discontinuity of the left atom \cite{hodgson2017interatomic,doi:10.1021/acs.jctc.0c01093}, $\Delta^{N=2}_\mathrm{L}$; Fig.~\ref{fig:Vary_delta_calculation} (bottom). (For this system $\Delta^{N=2}_\mathrm{L}=0.349$ a.u.) This plateau in the xc potential shrinks as $\delta \rightarrow 1$; the steps move towards the left atom until the plateau has zero width and the xc potential corresponds to the left atom with an additional electron ($\delta = 1$). 

Towards the right atom, at the point where the left atom's density meets the right's, there is a second change in the LEIP; see Fig.~\ref{fig:Vary_delta_calculation} (top). As predicted above, a plateau which elevates the xc potential in the region of the \textit{right atom} with magnitude $\Delta^{N=2}$ is present as $\delta \rightarrow 1$; see Figs.~\ref{fig:Vary_delta_calculation} (bottom) and \ref{fig:First_molecular_calculation} (bottom). 

In summary, when a finite but small fraction of an electron is added to this stretched molecule, there are \textit{two} plateaus in the xc potential -- one localized to the left atom with height $\Delta^{N_0}_\mathrm{L}$ and one localized to the right atom with height $\Delta^{N_0}$. At the edge of each plateau between the atoms there is a step. (Below I investigate the effect on the exact xc potential when the heights of these two plateaus are \textit{not} equal.) As the amount of additional electron tends to one, the left plateau (in the vicinity of the atom to which the additional electron localizes) is removed from the xc potential while the right plateau remains at a height of $\Delta^{N_0}$. Thus, within the ($N_0+1$)-electron system the right atom's xc potential is elevated by $\Delta^{N_0}$ yielding the interatomic step; see Fig.~\ref{fig:First_molecular_calculation} (bottom).

\section{An interatomic step to correctly distribute electrons} \label{sec:Curcial_step}

\begin{figure}[htbp]
  \centering
  \includegraphics[width=1.0\linewidth]{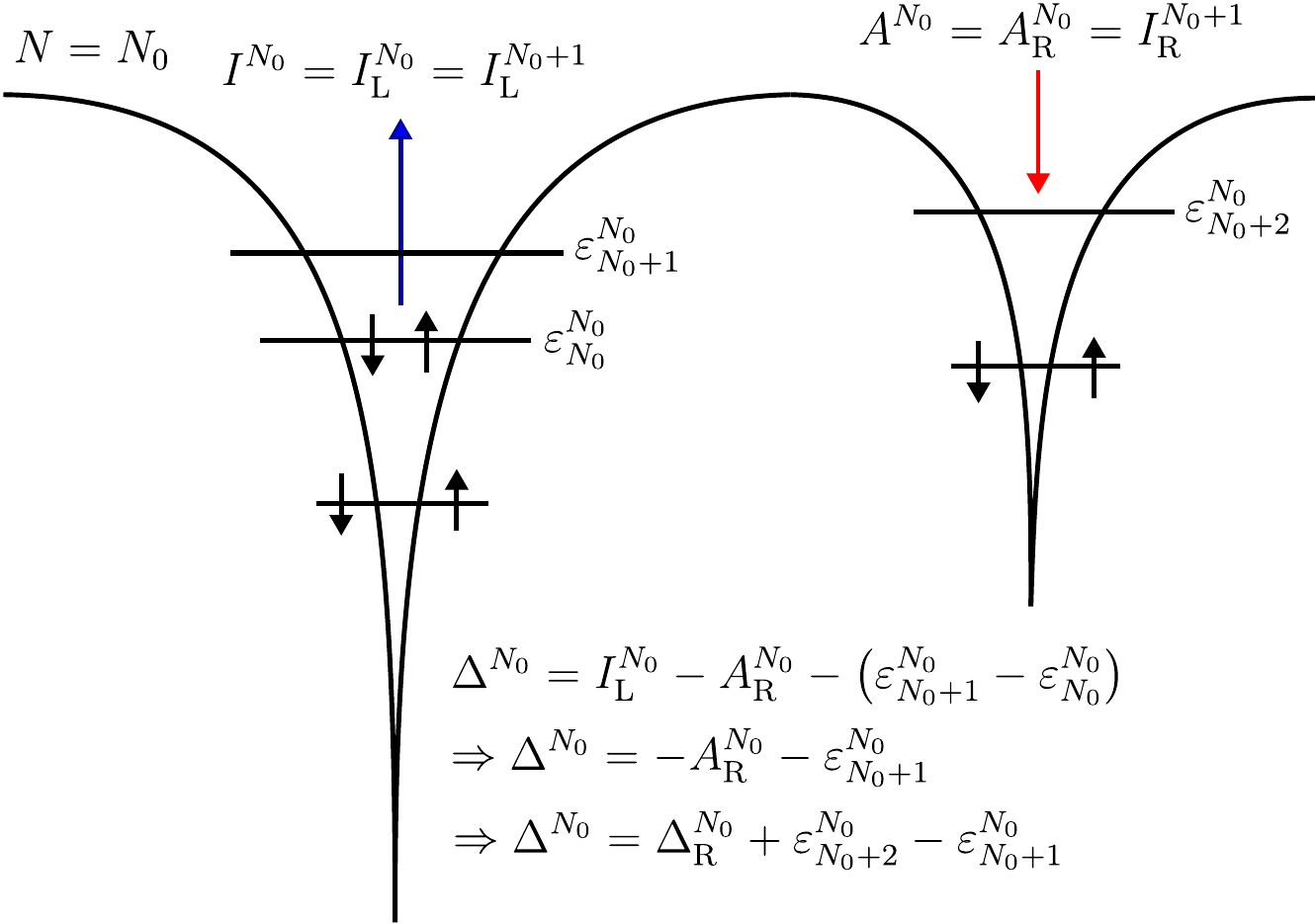}
\caption{Diagram of an $N_0$-electron stretched diatomic molecule where both atoms are closed shell \textit{and} the lowest unoccupied and highest occupied molecular orbitals are localized to the left atom but are \textit{not} degenerate.}
\label{fig:Second_molecule}
\end{figure}

Imagine an $N_0$-electron stretched diatomic molecule where the system is asymmetric (as above) and both atoms are closed shell; see Fig.~\ref{fig:Second_molecule}.

Consider the atoms to be so separated that the ionization energy of the molecule corresponds to the ionization energy of the isolated \textit{left} atom ($I^{N_0}=I^{N_0}_\mathrm{L}$) and the electron affinity of the whole molecule correspond to the affinity of the isolated \textit{right} atom ($A^{N_0}=A^{N_0}_\mathrm{R}$). Hence for this system $\Delta^{N_0} = I^{N_0}_\mathrm{L} - A^{N_0}_\mathrm{R} - \left( \varepsilon^{N_0}_{N_0+1} - \varepsilon^{N_0}_{N_0} \right ) = - A^{N_0}_\mathrm{R} -\varepsilon^{N_0}_{N_0+1}$. The HOMO and the LUMO are \textit{both} localized to the left atom but are not degenerate; see Fig.~\ref{fig:Second_molecule}. When an electron is added to this system it localizes to the right atom owing to the Coulomb and Pauli interaction which would be experienced if it were to localize to the left atom. Within the exact KS description if an electron were to be added to the exact KS potential of the $N_0$-electron system, it would spuriously localize to the left atom (within the many-body system it would localize in the right atom). Therefore, like for the Almbladh and von Barth case, an interatomic step in the ($N_0+1$)-electron is required to ensure that the KS single-particle density is equal to the many-body density by `reordering' the KS energies. 

When an infinitesimal amount of electron is added to this $N_0$-electron system the xc potential of the \textit{whole} system jumps by $\Delta^{N_0}$ (demonstrated above).

\begin{figure}[htbp]
  \centering
  \includegraphics[width=1.0\linewidth]{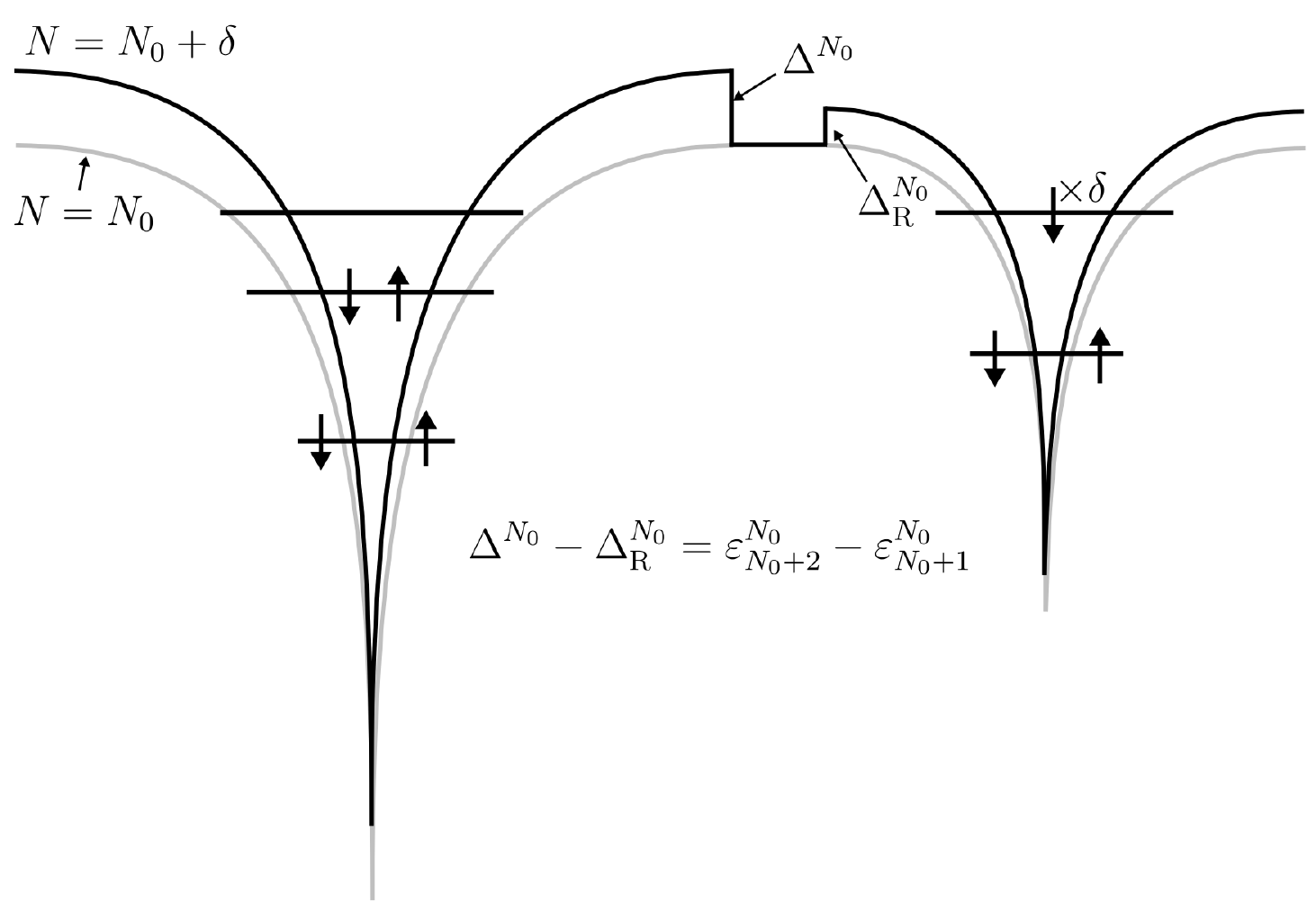}
\caption{Diagram of an ($N_0+\delta$)-electron stretched diatomic molecule with a closed-shell left atom and a partially occupied open-shell right atom within the exact KS description.}
\label{fig:Second_molecule_plus_delta}
\end{figure}

For small $\delta$, the addition electron completely localizes to the right atom which gives rise to a plateau in its xc potential the height of which corresponds to the xc derivative discontinuity of the right atom, $\Delta^{N_0}_\mathrm{R}$; see Fig.~\ref{fig:Second_molecule_plus_delta}. The left atom's xc potential is elevated by $\Delta^{N_0}$ for $0 < \delta \leq 1$ (shown above). Thus the height of the interatomic step within the ($N_0+\delta$)-electron system is $\Delta^{N_0} - \Delta^{N_0}_\mathrm{R} = - A^{N_0}_\mathrm{R} -\varepsilon^{N_0}_{N_0+1} - \Delta^{N_0}_\mathrm{R}$. For this system $\Delta^{N_0}_\mathrm{R} = -A^{N_0}_\mathrm{R}-\varepsilon^{N_0}_{N_0+2}$ as $\varepsilon^{N_0}_{N_0+2}$ is the lowest unoccupied \textit{atomic} KS energy of the right atom. Therefore $\Delta^{N_0} - \Delta^{N_0}_\mathrm{R} = \varepsilon^{N_0}_{N_0+2}-\varepsilon^{N_0}_{N_0+1}$; see Fig.~\ref{fig:Second_molecule_plus_delta}. Note that for this system the step has aligned the highest (partially) occupied KS energy of the right atom with the lowest \textit{unoccupied} KS energy of the left atom. 

When $\delta \rightarrow 1$ the right plateau is removed from the xc potential and the left plateau remains; the mechanism for which is described above. Hence, within the ($N_0+1$)-electron system the right atom is elevated by the plateau of height $\Delta^{N_0}$ and thus there is an interatomic step in the xc potential with height $S = \Delta^{N_0}$. 

\begin{figure}[htbp]
  \centering
  \includegraphics[width=1.0\linewidth]{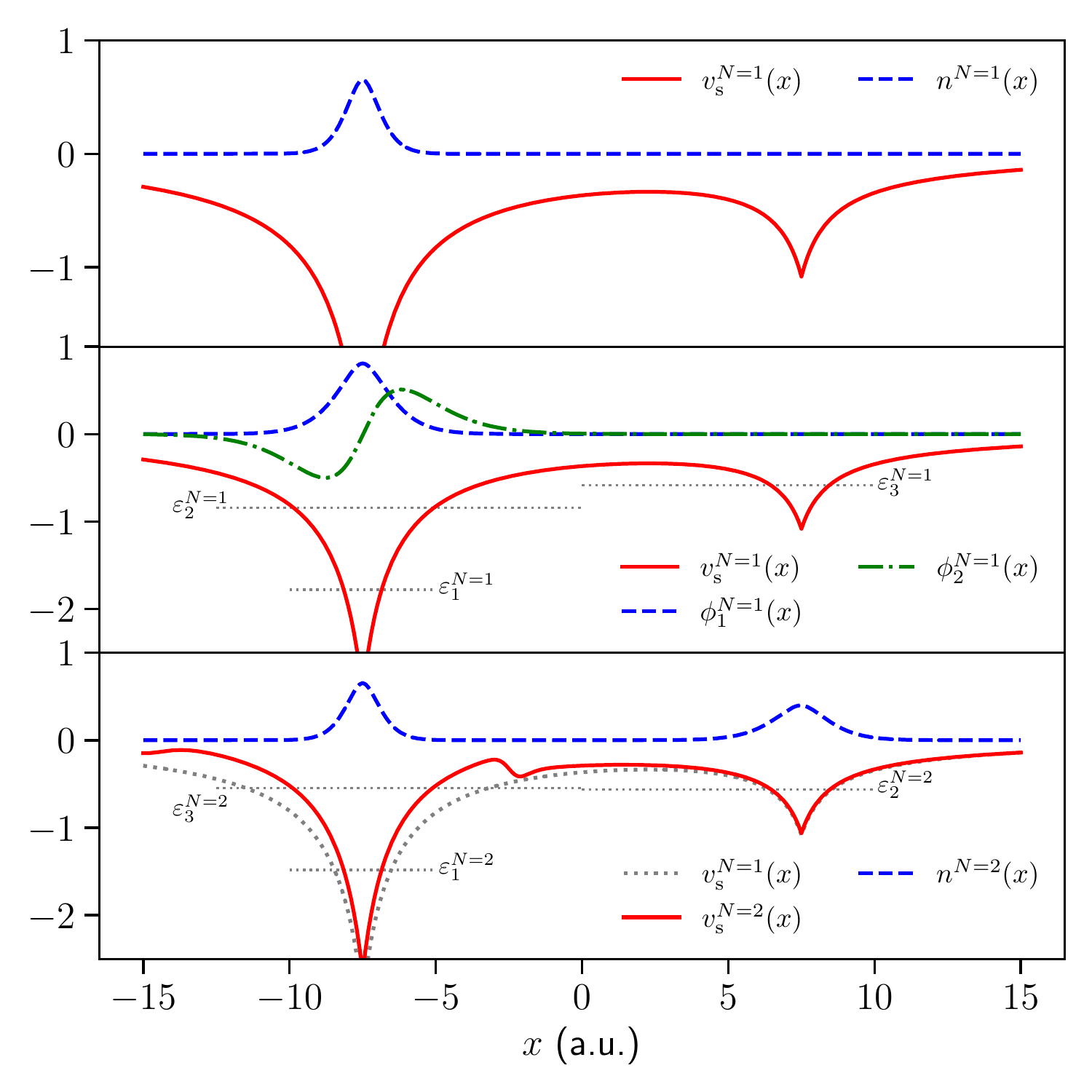}
\caption{(Top) The 1-electron density and the external potential. (Middle) The lowest unoccupied KS orbital, $\phi^{N=1}_2(x)$, and the highest occupied KS orbital, $\phi^{N=1}_1(x)$ of the 1-electron system. The corresponding single-particle energy levels are shown in gray. (Bottom) The many-body density of two same-spin electrons and the corresponding exact KS potential. The step elevates the xc potential in the vicinity of the left well in order to reorder the occupation of the KS states.}
\label{fig:second_molecule}
\end{figure}

For this system, as $A^{N_0}_\mathrm{R}=I^{N_0+1}_\mathrm{R}$, $\Delta^{N_0} = - I^{N_0+1}_\mathrm{R} - \varepsilon^{N_0}_{N_0+1}$. Within the ($N_0+1$)-electron system $I^{N_0+1}_\mathrm{R}=-\varepsilon^{N_0+1}_{N_0+1}$ owing to the IP theorem. Therefore, $\Delta^{N_0} = \varepsilon^{N_0+1}_{N_0+1}-\varepsilon^{N_0}_{N_0+1}$; see Fig.~\ref{fig:second_molecule} (bottom). Thus
\begin{equation} \label{Eq:SHCS_3}
    \lim_{d \rightarrow \infty} S^{N_0+1} = \Delta^{N_0} = \varepsilon^{N_0+1}_{N_0+1}-\varepsilon^{N_0}_{N_0+1}.
\end{equation}

I model a 1D stretched diatomic molecule which initially consists of 1 electron ($N_0=1$). The additional electron has the same spin as the initial electron so that the atoms are closed shell. $v_\mathrm{ext} (x) = -1/(|x-\frac{d}{2}| + 1) - 3/(|x+\frac{d}{2}| + 1)$ and $d=15$ a.u.; see Fig.~\ref{fig:second_molecule} (top). The left external potential well is deep enough that two same-spin fully non-interacting electrons in \textit{this external potential} both occupy the left well; see Fig~\ref{fig:second_molecule} (middle). However, within the triplet 2-electron many-body system one electron localizes in the left well and the other electron localizes in the opposite well owing to the Coulomb and Pauli interaction they would experience if they were to occupy the same well; see Fig.~\ref{fig:second_molecule} (bottom).

Equation~\eqref{Eq:SHCS_3} implies that the role of the step is to ensure that as the atoms are separated the difference between the highest occupied atomic KS energy of the right atom and the lowest unoccupied atomic KS energy of the left atom tends to zero. For any large but finite separation the highest occupied atomic KS energy of the right atom is slightly lower than the lowest unoccupied atomic KS energy of the left atom because the additional electron localized to the right atom has a small effect on the left atom's density. In practice this avoids the spurious occupation of the LUMO (for finite separations the inequality of Ref.~\onlinecite{PhysRevB.93.155146} must be obeyed by the step height); see Fig.~\ref{fig:second_molecule} (bottom). As I increase the separation of the external potential wells I find that this difference indeed tends to zero as predicting by Eq.~\eqref{Eq:SHCS_3}.

\subsection{Almbladh and von Barth case}

The Almbladh and von Barth case is similar to the that of Sec.~\ref{sec:Curcial_step} but both the HOMO and LUMO are localized to the now \textit{open-shell} left atom and the states are degenerate ($\varepsilon^{N_0}_{N_0}=\varepsilon^{N_0}_{N_0+1}$). Therefore $\Delta^{N_0} = I^{N_0}_\mathrm{L} - A^{N_0}_\mathrm{R}$; see Eq.~(\ref{Eq:DD}).

When an electron with the \textit{opposite} spin to the unpaired electron is added to this system within the many-body description, the additional electron localizes to the \textit{right} atom owing to the Coulomb interaction which it would experience if it tried to localize to the left atom. Note that within the exact KS system, despite the LUMO being localized to the left atom within the $N_0$-electron system, the additional KS electron must localize to the right atom in order for the single-particle KS density to equal the exact many-body electron density (as for the molecule of Sec.~\ref{sec:Curcial_step}). As such the highest occupied \textit{atomic} energies must be aligned within the ($N_0+1$)-electron KS system, which is ensured within the exact system by the interatomic step; see Eq.~(\ref{Eq:AvB}).

Within the ($N_0+\delta$)-electron system, the localization of the additional electron within the exact KS system is the same as described above; a plateau of height $\Delta_\mathrm{R}^{N_0}$ forms for the right atom (the atom to which the additional electron localizes) while a plateau of height $\Delta^{N_0}$ forms in the vicinity of the left atom. Similar to the molecule of Sec.~\ref{sec:Curcial_step}, $\Delta^{N_0} - \Delta^{N_0}_\mathrm{R} = \varepsilon^{N_0}_{N_0+2}-\varepsilon^{N_0}_{N_0+1}$, so in this case the step acts to align the highest (partially) occupied KS energy of the right atom with the highest occupied KS energy of the left atom, as required by Almbladh and von Barth's argument \cite{NATO85_AvB}. 

When $\delta=1$ the right plateau has gone and the left plateau is still present. For this stretched molecule $A^{N_0}_\mathrm{R}=I^{N_0+1}_\mathrm{R}$ and, because the additional electron localizes to the right atom $I^{N_0}_\mathrm{L}=I^{N_0+1}_\mathrm{L}$, where $I^{N_0+1}_\mathrm{L}$ is the ionization energy of the left atom and $I^{N_0+1}_\mathrm{R}$ is the ionization energy of the right atom within the ($N_0+1$)-electron system. Therefore $\Delta^{N_0} = I^{N_0+1}_\mathrm{L} - I^{N_0+1}_\mathrm{R}$. Thus, in accordance with Eq.~(\ref{Eq:AvB}), the step height is given by the difference between the left and right atoms' ionization energies:
\begin{equation} \label{Eq:SHCS_2}
    \lim_{d \rightarrow \infty} S^{N_0+1} = \Delta^{N_0} = I^{N_0+1}_\mathrm{L} - I^{N_0+1}_\mathrm{R}.
\end{equation}

\section{Conclusion}

I derived exact analytic expressions for the height of the spatial `step' that forms in the exact exchange-correlation (xc) potential between a variety of open- and closed-shell atoms within a stretched diatomic molecule via its relationship to the xc `derivative discontinuity'. I verified my results via exactly solvable 1D model systems. 

I consider the addition of an electron to an $N_0$-electron diatomic molecule in order to find an expression for the interatomic step in the ($N_0+1$)-electron system: When the amount of additional electron is infinitesimal, the \textit{whole} system's xc potential shifts discontinuously by a constant, $\Delta^{N_0}$ (the xc derivative discontinuity of the $N_0$-electron system). For sufficiently large separation between the atoms of this molecule, the additional electron completely localizes to one of the atoms, say the left. This left atom experiences a local shift (a plateau) by its xc derivative discontinuity, $\Delta^{N_0}_\mathrm{L}$, which may differ to that of the whole molecule, $\Delta^{N_0}$. In conjunction a plateau with height $\Delta^{N_0}$ localizes to the right atom; depicted in Fig.~\ref{fig:small_delta} in Sec.~\ref{sec:infinitesimal_additional_e}.

As the amount of additional electron is increased to 1, the plateau localized to the left atom, to which the additional electron localizes, shrinks while the plateau localized to the right atom has a fixed height of $\Delta^{N_0}$; shown in Sec.~\ref{sec:infinitesimal_additional_e}. As a result, within the ($N_0+1$)-electron system the right atom's xc potential is elevated by $\Delta^{N_0}$ which means a step with height $\Delta^{N_0}$ is present in the xc potential \textit{between} the atoms. 

I present three distinct example diatomic molecules that consist of a variety of open- and closed-shell atoms. In each case the height of the interatomic step, $S=\Delta^{N_0}$. Two of my example systems require a step in the xc potential to `reorder' the KS single-particle energies so that the KS density equals the many-body density (one of which is the well-known thought experiment of Almbladh and von Barth \cite{NATO85_AvB}). Without the step, the KS electron density would experience a large delocalization error \cite{hellgren2019strong}. I demonstrate that the required reordering of the KS single-particle states is achieved completely by the interatomic step of height $\Delta^{N_0}$; see Eq.~\eqref{Eq:SHCS_3}, which is verified numerically in Fig.~\ref{fig:second_molecule} (bottom). I show that in the case of two open-shell atoms $S=\Delta^{N_0}$ is consistent with Almbladh and von Barth's result; see Eq.~\eqref{Eq:SHCS_2}. 

An accurate description of the step structures in the xc potential is needed to correctly distribute electron density throughout a molecular structure, describe the dissociation of atoms, bonding lengths, ground-state energies, and excitation energies \cite{doi:10.1021/acs.jctc.0c01093,kocak2021charge}. For example, existing approximate functionals that omit these step features cannot be employed to reliably calculate optical absorption properties for charge-transfer donor-acceptors \cite{PhysRevA.85.022514,PhysRevA.88.052507,hellgren2019strong,doi:10.1021/acs.jctc.0c01093}. My expressions for the step height could be employed to aid the development of advanced approximations to the xc energy and corresponding xc potential in order to yield more reliable KS calculations and offer a path to the computationally efficient calculation of excitation energies within quantum chemistry alleviating the reliance on hybrid density functionals that carry a large computational cost. 

\begin{acknowledgments}
I thank the University of York for computational resources as well as Rex Godby and Nikitas Gidopoulos for helpful discussions.
\end{acknowledgments}

\bibliographystyle{apsrev4-2}

\end{document}